\documentclass[a4paper,reprint,aps,pre,floatfix,twocolumn,superscriptaddress,showpacs,nofootinbib]{revtex4-1}
  
\usepackage[pdftex]{graphicx}
\usepackage{latexsym,amsmath,amssymb}
\usepackage[english]{babel}
\usepackage[pdftex]{color}
\usepackage{hyperref}
\usepackage{microtype}
\usepackage{comment}
\usepackage[normalem]{ulem}

\newcommand{\av}[1]{\langle #1 \rangle}
\newcommand{\km}{k_\textrm{max}}

\newcommand{\lrangle}[1]{\langle{#1}\rangle}

\newcommand{\lbkout}[1]{\lambda_{#1}^\mathrm{out}}
\newcommand{\lbkin}[1]{\lambda_{#1}^\mathrm{in}}

\newcommand{\FigPath}{./}

\begin{document}

\title{Collective versus hub activation of epidemic phases on networks}

\author{Silvio C. Ferreira} \email{silviojr@ufv.br}

\author{Renan S. Sander}

\affiliation{Departamento de F\'{\i}sica, Universidade Federal de
  Vi\c{c}osa, 36570-000, Vi\c{c}osa, MG, Brazil}

\author{Romualdo Pastor-Satorras}

\affiliation{Departament de F\'{\i}sica, Universitat Polit\`ecnica de
  Catalunya, Campus Nord B4, 08034 Barcelona, Spain}

\begin{abstract}
  We consider a general criterion to discern the nature of the threshold
  in epidemic models on scale-free (SF) networks. Comparing the epidemic
  lifespan of the nodes with largest degrees with the infection time
  between them, we propose a general dual scenario, in which the
  epidemic transition is either ruled by a hub activation process,
  leading to a null threshold in the thermodynamic limit, or given by a
  collective activation process, corresponding to a standard phase
  transition with a finite threshold.  We validate the proposed
  criterion applying it to different epidemic models, with waning
  immunity or heterogeneous infection rates in both synthetic and real
  SF networks. {In particular, a waning immunity, irrespective of its
  strength, leads to collective activation with finite threshold in
  scale-free networks with large exponent, at odds with canonical
  theoretical approaches.}
\end{abstract}


\maketitle
\section{Introduction}

The study of epidemic spreading \cite{Pastor-Satorras:2014aa} in complex
topologies is one of the cornerstones of modern network science
\cite{Newman10}, with applications in the spread of influence, opinions
and other social phenomena
\cite{David:2010:NCM:1805895,RevModPhys.81.591}. Of particular interest
is the theoretical understanding of epidemic models in scale-free (SF)
networks \cite{Barabasi:1999}, in which the probability $P(k)$ (degree
distribution) that a node is connected to $k$ others (has degree $k$)
exhibits heavy tails of the form $P(k) \sim k^{-\gamma}$.  This interest
is motivated by the possible effects that a heterogeneous topology might
have on the location of the epidemic threshold $\lambda_c$, for some
control parameter $\lambda$, signaling a phase transition separating a
healthy, disease-free phase, from an infected phase, in which the
epidemics can thrive \cite{Pastor-Satorras:2014aa}.

For epidemics leading to a steady (endemic) state, the main object of
interest has been the susceptible-infected-susceptible (SIS) model,
which is defined as follows \cite{anderson92}: Individuals, represented
as nodes in the network, can assume two different states, susceptible
($S$) or healthy, and infected ($I$), and are capable to transmit the
disease. Infected individuals recover and become spontaneously
susceptible again with a rate $\beta$ that can be taken equal to
$1$. Transmission of the disease is effected by a rate to transmit the
disease through an edge connecting an infected to a susceptible node
equal to a constant $\lambda$.  After a considerable theoretical effort,
it has been shown that the behavior of the SIS model in uncorrelated
\cite{Newman10} SF networks is far from trivial
\cite{Pastor01,Boguna02,Chakrabarti08,chatterjee2009,Castellano10,Ferreira12,
  Castellano2012,boguna2013nature, Mountford2013,Goltsev12}.  Two
competing theories were initially proposed to account for the SIS
epidemic threshold.  Heterogeneous mean-field (HMF) theories
\cite{Pastor-Satorras:2014aa}, neglecting both dynamical and topological
correlations, provide a threshold
$\lambda_c^\mathrm{HMF} = \av{k}/\av{k^2}$ \cite{Pastor01b,Boguna02},
which tends to zero in the thermodynamic limit for $\gamma \leq 3$, and
is finite for $\gamma > 3$.  Quenched mean-field (QMF) theory
\cite{Chakrabarti08}, including the full network structure through its
adjacency matrix $A_{ij}$ \cite{Newman10}, but still neglecting
dynamical correlations, predicts instead a threshold
$\lambda_c^\mathrm{QMF} \simeq 1 /\Lambda_m$, where $\Lambda_m$ is the
largest eigenvalue of the adjacency matrix. The scaling form of this
threshold is given by~\cite{Castellano10,Chung03}
$\lambda_c^\mathrm{QMF} \simeq 1 /\sqrt{k_\mathrm{max}}$ for
$\gamma>5/2$, where $k_\mathrm{max}$ is the maximum degree in the
network, while for $\gamma<5/2$ it yields
$\lambda_c^\mathrm{QMF} \simeq \av{k}/\av{k^2}$ in agreement with HMF
theory.  Numerical simulations
\cite{Castellano10,Ferreira12,boguna2013nature} indicate that QMF is
qualitatively correct in SF networks, implying that the epidemic
threshold vanishes in the thermodynamic limit when $k_\mathrm{max}$
diverges, irrespective of the degree exponent $\gamma$.

The origin of this null threshold has been physically interpreted in
Ref.~\cite{boguna2013nature} taking explicitly into account the
interplay between the lifetime of a hub of degree $k$,
$\tau^\mathrm{rec}_k$, and the time scale $\tau^\mathrm{inf}_{k, k'}$
with which an infected hub of degree $k$ infects a susceptible hub of
degree $k'$. The fact that the lifetime $\tau^\mathrm{rec}_k$ is
diverging with degree $k$ faster than $\tau^\mathrm{inf}_{k, k'}$ for
any value of $\lambda$ is the ultimate cause of the null epidemic
threshold in the SIS model
\cite{boguna2013nature,chatterjee2009,Mountford2013}.  However, other
epidemic and dynamical models on SF networks, in particular the contact
process (CP)\cite{Castellano:2006}, defined by an infection rate
inversely proportional to the degree of the infected node, $\lambda/k$,
possess a finite threshold which can be better captured in terms of a
degree-based HMF theory \cite{cpquenched,Mata14}. This observation
claims for an understanding of the mechanisms ruling epidemic
transitions, regarding in particular the conditions under which the
threshold is either constant or vanishing.

The results of Ref.~\cite{boguna2013nature}, while important, are
strictly applied to the SIS process, and thus a general formalism,
adapted to more complex and realistic epidemic models with a steady
state, is still lacking.  Here we determine the recovery
$\tau^\mathrm{rec}$ and infection $\tau^\mathrm{inf}$ time scales of
hubs for generic epidemic models including waning immunity and arbitrary
edge-dependent infection rates and, building on these results, we
propose a classification of endemic epidemic transitions on networks:
When $\tau^\mathrm{rec} \gg \tau^\mathrm{inf}$, a scenario of local hub
activation with mutual hub reinfection is at work
\cite{boguna2013nature}, leading to $\lambda_c\rightarrow 0$ when the
recovery times diverges in the thermodynamic limit. In such scenario,
QMF theories are expected to be qualitatively correct. On the other
hand, for {$\tau^\mathrm{rec} \leq \tau^\mathrm{inf}$}, mutual hub
reinfection cannot take place, and an endemic state is possible only
through a collective activation of the whole network in a standard phase
transition occurring at a finite threshold. In this second scenario, HMF
theories should be correct.

We present evidence for this scenario analyzing the
susceptible-infected-removed-susceptible (SIRS) model~\cite{anderson92},
an extension of the SIS model allowing for a waning immunity of
nodes. While it has be shown that SIRS is equivalent to SIS dynamics in
the framework of standard mean-field theories
\cite{ForestFireSatorras09}, here we show that the effect of waning
immunity is to induce a finite threshold in SF networks for $\gamma>3$,
at odds with QMF and qualitatively described by HMF theory. For sake of
generality, the theory is also applied to other epidemic models without
immunity, namely, the CP~\cite{Castellano:2006} and the generalized SIS
model proposed by Karsai, Juh\'asz, and Igl\'oi (KJI) \cite{Karsai} with
weighted infection rates.

Our paper is organized as follows: We develop the theory for the
interplay between of hub lifetime and mutual hub infection time in
Section~\ref{sec:theory} and corroborate the theory with simulations on
synthetic and real SF networks in sections~\ref{sec:simu} and
\ref{sec:real}, respectively. We briefly summarize our conclusions and
prospects in section~\ref{sec:conclu}.  Three appendices complement the
paper: HMF and QMF theories for investigated models are presented in
appendices \ref{app:KJI} and \ref{app:SIRS} while simulation methods are
presented in appendix \ref{app:simu}.

\section{Theory}
\label{sec:theory}

\subsection{The generalized epidemic model}
\label{sec:gener-epid-model}

\begin{figure}[htb]
\centering
\includegraphics[width=0.99\linewidth]{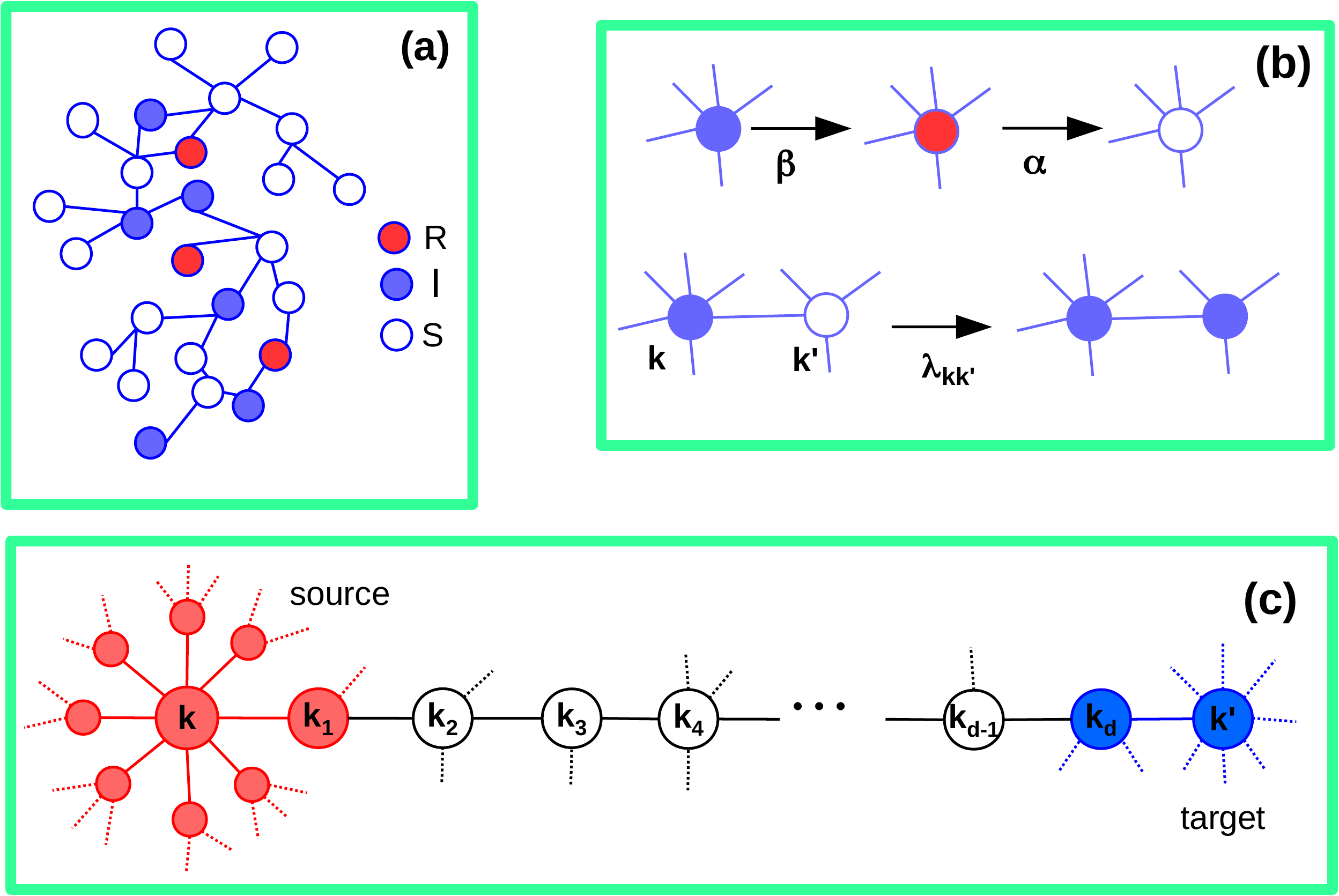}
\caption{(color online) Generic epidemic model in complex
  networks. {(a)} Nodes in the network can be either healthy or
  susceptible ($S$), infected ($I$), and immune or recovered
  ($R$). {(b)} Infected individuals recover spontaneously, $I \to R$,
  with rate $\beta$.  Recovered individuals become again spontaneously
  susceptible, $R \to S$, with a rate $\alpha$. Heterogeneous
  transmission is implemented by infected nodes of degree $k$
  transmitting the disease to adjacent susceptible nodes of degree $k'$
  with rate $\lambda_{k,k'}$. {(c)} Schematic representation of chain of
  $d$ vertices of arbitrary degrees $k_1,k_2,\ldots,k_d$, connecting two
  hubs of degree $k$ and $k'$. The infection starting in the center of
  the leftmost star (red) aims to reach the rightmost star (blue). Stubs
  as dashed lines represent edges that can transmit but cannot receive
  the infection.}
\label{fig:chain_trasm}
\end{figure}

To develop our theory we study a generalized epidemic model on a network
where each vertex can be either healthy or susceptible ($S$), infected
($I$), and immune or recovered ($R$). Infected individuals recover
spontaneously, $I \to R$, with rate $\beta$. Recovered individuals
become again spontaneously susceptible (waning immunity) $R \to S$, with
a rate $\alpha$. Infected nodes of degree $k$ transmit the disease to
each adjacent susceptible node of degree $k'$ with an heterogeneous
infection rate $\lambda_{k,k'}$. See Figs.~\ref{fig:chain_trasm}(a) and
\ref{fig:chain_trasm}(b) for a graphical description of the model. From
this generalized epidemic model, classical ones can be recovered: The
SIS model ($\alpha\to\infty$, $\lambda_{k,k'} = \lambda$); the
CP~\cite{Castellano:2006} model ($\alpha\to\infty$,
$\lambda_{k,k'} = \lambda/k$); the SIRS model ($\alpha$ finite,
$\lambda_{k,k'} = \lambda$); the KJI~\cite{Karsai} model
($\alpha\to\infty$, $\lambda_{k,k'} = \lambda/(k k')^\theta$), etc.

To be used later, we define the average infection rate produced,
$\lbkout{k}$, and received, $\lbkin{k}$, by a vertex of degree $k$ as
\begin{equation}
  \label{eq:1}
  \lbkout{k}=\sum_{k'}\lambda_{kk'}P(k'|k), \quad
  \lbkin{k}=\sum_{k'}\lambda_{k'k}P(k'|k) ,
\end{equation}
respectively, where $P(k'|k)$
is the probability that a vertex of degree $k$ is connected to vertex of
degree $k'$~\cite{alexei}.

\subsection{Hub lifetime}
\label{sec:hub-lifetime}

We focus in the first place on the hub lifetime $\tau^\mathrm{rec}_k$,
which is defined as the average time that a hub of degree $k$, starting
from a configuration with a single infected node, takes to reach a
configuration in which the hub and its nearest neighbors are all
susceptible.  To estimate this quantity, we approximate the dynamics of
a hub of degree $k$ by that of a star-like graph, composed by a center
connected to $k$ nodes (leaves) of arbitrary degree, the red nodes in
Fig.~\ref{fig:chain_trasm}(c), where we neglect infection coming from
outside the star. For analytical tractability, we consider a simplified
generic epidemic process with two states ($S$, $I$) in the leaves and
three states ($S$, $I$, $R$) in the center. The rate of infection along
an edge from a leaf to the center is approximated by $\lbkin{k}$ and
from the center to a leaf by $\lbkout{k}$, see Eq.~(\ref{eq:1}); that
is, we consider the effect of the leaves as an average over their
possible degree values $k'$, weighted with the probability $P(k' |
k)$. The transitions $I \to S$ (leaves), $I \to R$ and $R \to S$
(center) have constant rates $\beta$, $\beta$, and $\alpha$,
respectively. The lifespan for this dynamics is larger than in the real
model, where leaves can also assume the $R$ state, and so it provides an
upper bound for the true $\tau^\mathrm{rec}_k$.

We approximate this dynamics in a star as follows: 

i) At $t=0$ the center is infected and all leaves are susceptible.

ii) At time $t_1=1/\beta$ the center becomes recovered and $n_1$
leaves are infected with probability
\begin{equation}
P_1(n_1|k)=\binom{k}{n_1} p_1^{n_1}(1-p_1)^{k-n_1},
\end{equation}
where $p_1=1-\exp(-\lbkout{k}/\beta)$ is the probability that each
leaf was infected by the center in the interval $t<t_1$.

iii) At time $t=t_1+t_2$, where $t_2$ has a distribution
$\rho_2(t_2) = \alpha \exp(-\alpha t_2)$, the center becomes susceptible
and $n_2$ leaves remain infected with probability
\begin{equation}
P_2(n_2|n_1)=\binom{n_1}{n_2} p_2^{n_2}(1-p_2)^{n_1-n_2},
\end{equation}
where $p_2=\exp(-\beta t_2)$ is the probability that each active leaf
remains infected for a time $t_2$.

iv) At time $t=t_1+t_2+t_3$, where $t_3=1/\beta$, all $n_2$ leaves infected at
time $t_1+t_2$ become (synchronously) susceptible and the center is infected
again with probability
\begin{equation}
P_3(n_2)= 1-(1-p_3)^{n_2},
\end{equation}
where $p_3=1-\exp(-\lbkin{k}/\beta)$ is the probability that each leaf sent the
infection to the center during a time $t_3$.

Steps ii) and iv) are essentially a generalization of the approximation
for SIS dynamics on stars in Ref.~\cite{boguna2013nature}, in which
stochasticity of infective time and multiple infections of the leaves
are neglected, while step ii) does not involve approximations. Treating
step ii) stochastically is essential since the rare events in which only
a few infected leaves survive cannot be neglected. The probability that
the star returns to its initial state in one step with interevent times
$t_1,~t_2$ and $t_3$ is
\begin{equation}
q_k(t_2) = \sum_{n_1=1}^k P_1(n_1|k)\sum_{n_2=1}^{n_1} P_2(n_2|n_1)P_3(n_2)
\end{equation}
Averaging $q_k(t_2)$ over $\rho_2(t_2)$ we finally have
\begin{equation}
  Q_k =  1-\alpha\int_0^\infty 
  e^{-\alpha t_2}    [1-e^{-\beta t_2} A]^k  dt_2,
  \label{eq:2}
\end{equation}
with $A=(1-e^{-\lbkout{k}/\beta})(1-e^{-\lbkin{k}/\beta})$. Now, the
probability that this dynamics survives for $s$ steps of average
duration $\tau_0 = t_1 + \av{t_2} + t_3 = 2/\beta + 1/\alpha$ is
$P(s)=Q_k^{s-1}(1-Q_k)$ and the average number of survival steps is
$\lrangle{s}=\sum_{s=1}^\infty sP(s) = 1/(1-Q_k)$. We thus obtain the
final result
\begin{equation}
\tau^\mathrm{rec}_k \equiv \tau_0 \av{s} = \frac{\tau_0}{1-Q_k}.
\label{eq:tauQk}
\end{equation}

Considering the absence of waning immunity inserting in Eq.~(\ref{eq:2})
the limit
$\lim_{\alpha\rightarrow\infty} \alpha e^{-\alpha t} = \delta(t)$, the
Dirac delta function, and assuming
$\lbkin{k}/\beta, \lbkout{k}/\beta \ll 1$, we obtain
\begin{equation} 
  \tau^\mathrm{rec}_k  \sim \exp( k \lbkin{k} \lbkout{k}/\beta^2).
  \label{eq:taukSIS}
\end{equation}

In the case of $\alpha$ finite, Eq.~\eqref{eq:2} becomes, after the
change of variable $u = A e^{-\beta t},$
\begin{equation}
  Q_k  = 1-\frac{\alpha}{\beta
    A^{\alpha/\beta}} %
\int_0^A u^{a-1} (1-u)^k \,du.
\label{eq:Q1}
\end{equation}
We can estimate the behavior of $Q_k$ in the limit of large $k$ 
using
\begin{equation}
  (1-u)^k = \exp\left[k \ln(1-u)\right] \simeq e^{-u k},
\end{equation}
valid for $0<u\ll 1$, and
\begin{equation}
  \int_0^A u^{a-1} e^{-u k}  \,du \simeq \int_0^\infty u^{a-1} e^{-u k}
  \,du = k^{-a} \Gamma(a),
\end{equation}
where $\Gamma(z)$ is the gamma function \cite{abramovitz}, and in which
the extension of the integral's upper limit to infinity is valid for
$kA$ large. Thus, from Eq.~\eqref{eq:Q1}, we obtain
\begin{equation}
  Q_k\simeq 1-\frac{\alpha}{\beta
    A^{\alpha/\beta}}\Gamma\left(\frac{\alpha}{\beta}\right)k^{-\alpha/\beta}.  
\end{equation}
From here and Eq.~\eqref{eq:tauQk}, it follows  in the
limit of large $k$
\begin{equation} 
  \tau_k^\mathrm{rec}  \sim k^{\alpha/\beta} \left[
    (1-e^{-\lbkout{k}/\beta})(1-e^{-\lbkin{k}/\beta})\right]^{\alpha/\beta}. 
  \label{eq:taukSIRS}
\end{equation}

\subsection{Hub mutual infection time}

To estimate the infection time $\tau^\mathrm{inf}_{k, k'}$, we consider
two stars of degree $k$ (the source $i=0$) and $k'$ (the target at
$i=d+1$), connected through a path of $d$ vertices ($i=1,2,\ldots,d$) of
arbitrary degree, see Fig.~\ref{fig:chain_trasm}.
The following hypothesis are used in the derivation for the case
$\tau^\mathrm{rec}_k\gg 1/\lbkout{k}$:

{i}) The vertex with degree $k$ on the left ($i=0$) is never recovered
and transmits the infection to its nearest neighbor at $i=1$ at an
average rate $\lbkout{k}$ and, in an average, a new epidemic outbreak is
started at $i=1$ each $1/\lbkout{k}$ time units.
	
{ii}) We assume that both $\lambda_{kk'}/\beta\ll 1$ and
$\lambda_{kk'}/\alpha \ll 1$ and consider only the epidemic routes where
an infected vertex always transmit the infection to its right
neighbor before it becomes recovered or
susceptible~\cite{boguna2013nature}. Additionally, we assume the average
transmission rate for all edges inside the chain $(i=1,\ldots,d-1)$
$\bar{\lambda}=\sum_{k'}\lbkout{k}P(k)$ that leads to the average
probability of transmission per edge given by
$\bar{q}=\bar{\lambda}/(\bar{\lambda}+\beta)$.
	
{iii}) The rate at which an infected vertex at $i=d$ transmits the
infection to the rightmost hub of degree $k'$ at $i=d+1$ is approximated
by $\lbkin{k'}$. The average probability to transmit do $i=d+1$ before
recovering of $i=d$ is, therefore,
$\bar{q}_{k'}=\lbkin{k'}/(\lbkin{k'}+\beta)$.
	
{iv}) The probability that an infection started at $i=1$ reaches the
rightmost hub ($i=d+1$) under these assumptions is given by
$\bar{q}^{d-2}\bar{q}_{k'}$ and, consequently, the transmission rate is
$\lbkout{k} \bar{q}^{d-2}\bar{q}_{k'}$.
	
{v}) For small-world networks with $N$ vertices, the average distance
between nodes of degrees $k$ and $k$' is~\cite{Holyst}
\begin{equation}
  d=1+\ln(N\lrangle{k}/k k')/\ln \kappa,
\label{eq:3}
\end{equation}
where $\kappa=\lrangle{k^2}/\lrangle{k}-1$, resulting in an upper bound
for the infection time (inverse of the rate) given by
\begin{equation}
  \tau^{\mathrm{inf}}_{k, k'} \lesssim \tau_{k k'} =
  \frac{\bar{q}_{k'}}{\lbkout{k}\bar{q}} 
  \left(\frac{N\lrangle{k}}{k k'}\right)^{b(\bar{\lambda})}, 
 \label{eq:bound}
\end{equation}
where $b(\bar{\lambda})={\ln(1+{\beta}/{\bar{\lambda}})}/{\ln \kappa}$.

For the case $\tau^\mathrm{rec}_k< 1/\lbkout{k}$, the leftmost star
recovers before produce the first outbreak in $i=1$ and thus the
infection time becomes infinite.

\section{Analysis of epidemic models on synthetic scale-free networks}
\label{sec:simu}

In this Section we present the analysis of different epidemic models on
SF networks characterized by a degree distribution
$P(k)\sim k^{-\gamma}$, that in a network of finite size $N$ extends up
to a maximum degree \cite{mariancutofss}
\begin{equation}
  k_\mathrm{max}(N) \sim \left\{
    \begin{array}{cl}
      N^{1/2} & \;\mathrm{for}    \;\gamma<3 \\
      N^{1/(\gamma-1)} & \;\mathrm{for}    \;\gamma>3 \\
    \end{array}
  \right. .
  \label{eq:5}
\end{equation}
In these networks, we have
\begin{equation}
  \kappa \sim \left\{
    \begin{array}{cl}
      k_\mathrm{max}^{3-\gamma} \simeq N^{(3-\gamma)/2} & \;\mathrm{for}
                                                          \;\gamma<3 \\
      \mathrm{const.} &  \;\mathrm{for}
                       \;\gamma>3 
    \end{array}
  \right.
  \label{eq:4}
\end{equation}
We focus in particular in uncorrelated networks, with
$P(k' | k) = k' P(k')/\av{k}$ \cite{Dorogovtsev:2002}, as generated by
the uncorrelated configuration (UCM) model \cite{Catanzaro05}.

\subsection{SIS model}

The SIS model is defined by $\alpha = 0$ and $\lambda_{k,k'} = \lambda$,
independent of $k$ and $k'$, yielding
$\lbkin{k} = \lbkout{k} = \lambda$. This values imply, from
Eq.~(\ref{eq:taukSIS}),
$\tau_k^\mathrm{rec, SIS} \sim \exp(\lambda^2 k / \beta)$. On the other
hand, from Eq.~(\ref{eq:bound}), and using Eq.~(\ref{eq:4}), we obtain a
$\tau_k^\mathrm{inf, SIS}$ that is constant for $\gamma<3$ ($b \to 0$),
while for $\gamma>3$ ($b \to \mathrm{const}$) it shows an algebraic
increase, that, for the largest hubs with degree $k_\mathrm{max}$ given
by Eq.~(\ref{eq:5}) takes the form
$\tau^\mathrm{inf, SIS} \sim N^{\frac{\gamma-3}{\gamma-1}b(\lambda)}$.
Both expressions have been confirmed by means of numerical simulations
of the SIS model in Ref.~\cite{boguna2013nature}. We have thus that, for
the SIS model, $\tau_k^\mathrm{rec, SIS} \gg \tau^\mathrm{inf, SIS}$,
i.e. the hubs survive for much longer times than are needed for a hub to
reinfect another, and therefore it is plausible a scenario in which the
transition is ruled by a hub activation dynamics. This possibility is
substantiated by an analysis of the SIS dynamics within a theory taking
into account, at a mean-field level, the dynamics of hub recovery and
mutual reinfection, leading to a vanishing epidemic threshold scaling
with network size in qualitative agreement with the predictions of QMF
theory \cite{boguna2013nature}

\subsection{Contact process}

In the case of the CP~\cite{Castellano:2006}, we have $\alpha=\infty$
and $\lambda_{k,k'}=\lambda/k$ implying $\lbkin{k} =\lambda/\lrangle{k}$
and $\lbkout{k} = \lambda/k$, and thus, from Eq.~(\ref{eq:taukSIS}),
$\tau_k^\mathrm{rec,CP} \sim \mathrm{const}$. On the other hand,
$1/\lbkout{k}$, and thus $\tau^{\mathrm{inf,CP}}$, diverges implying
that, for any value of $\gamma$,
$\tau^\mathrm{rec,CP} \ll \tau^{\mathrm{inf,CP}}$.  This result
indicates that it is impossible to have a scenario in which the
transition is driven by the successive activation and reactivation of
hubs, and that the associated epidemic transition must be given by
collective phenomenon, involving the activation of the whole
network. This collective transition is consistent with the finite
threshold numerically observed in the CP on SF networks
\cite{cpquenched,Mata14}, in agreement with HMF
predictions\cite{Castellano:2006,Mata14}.  Interestingly in the case of
the CP, the QMF prediction~\cite{Gomez10} coincides with the HMF
theory~\cite{Castellano:2006}, $\lambda_c=1$, indicating a threshold
completely independent of the network structure. This theoretical
prediction it is not fully observed in numerical simulations, which show
a constant threshold but that is modulated by network heterogeneity
\cite{Castellano:2006,cpquenched,Mata14}

\subsection{KJI model}

The KJI model is defined by $\alpha \to \infty$ and a heterogeneous
infection rate $\lambda_{k, k'} = \lambda/ (k k')^\theta$, with
$0 \leq \theta \leq 1$; that is, the infection power decreases with the
degree of both the infected and susceptible nodes connected by the
corresponding edge.

Simple HMF theory \cite{Karsai} (see Appendix \ref{app:KJI}) predicts a
threshold in uncorrelated networks
\begin{equation}
  \lambda_c^\mathrm{HMF,KJI} = \frac{\av{k}}{\av{k^{2(1-\theta)}}},
  \label{eq:7}
\end{equation}
which takes a finite value for $\gamma > 3-2\theta$, and in particular
for $\gamma>3$ and any $\theta>0$. On the other hand, QMF theory (see
Appendix~\ref{app:KJI}) predicts 
\begin{equation}
  \lambda_c^\mathrm{QMF,KJI} = \frac{1}{\Lambda_m^D},
  \label{eq:8}
\end{equation}
where $\Lambda_m^D$ is the largest eigenvalue of the matrix
$D_{ij} = A_{ij}/(k_i k_j)^\theta$, $A_{ij}$ being the adjacency
matrix. This largest eigenvalue (see Appendix~\ref{app:KJI}) increases
with network size for $\theta<1/2$ irrespective of $\gamma$.  Therefore
the QMF prediction is a vanishing threshold for $\theta <1/2$, and a
finite one otherwise.

\begin{figure}[t]
 \includegraphics[width=0.8\linewidth]{\FigPath/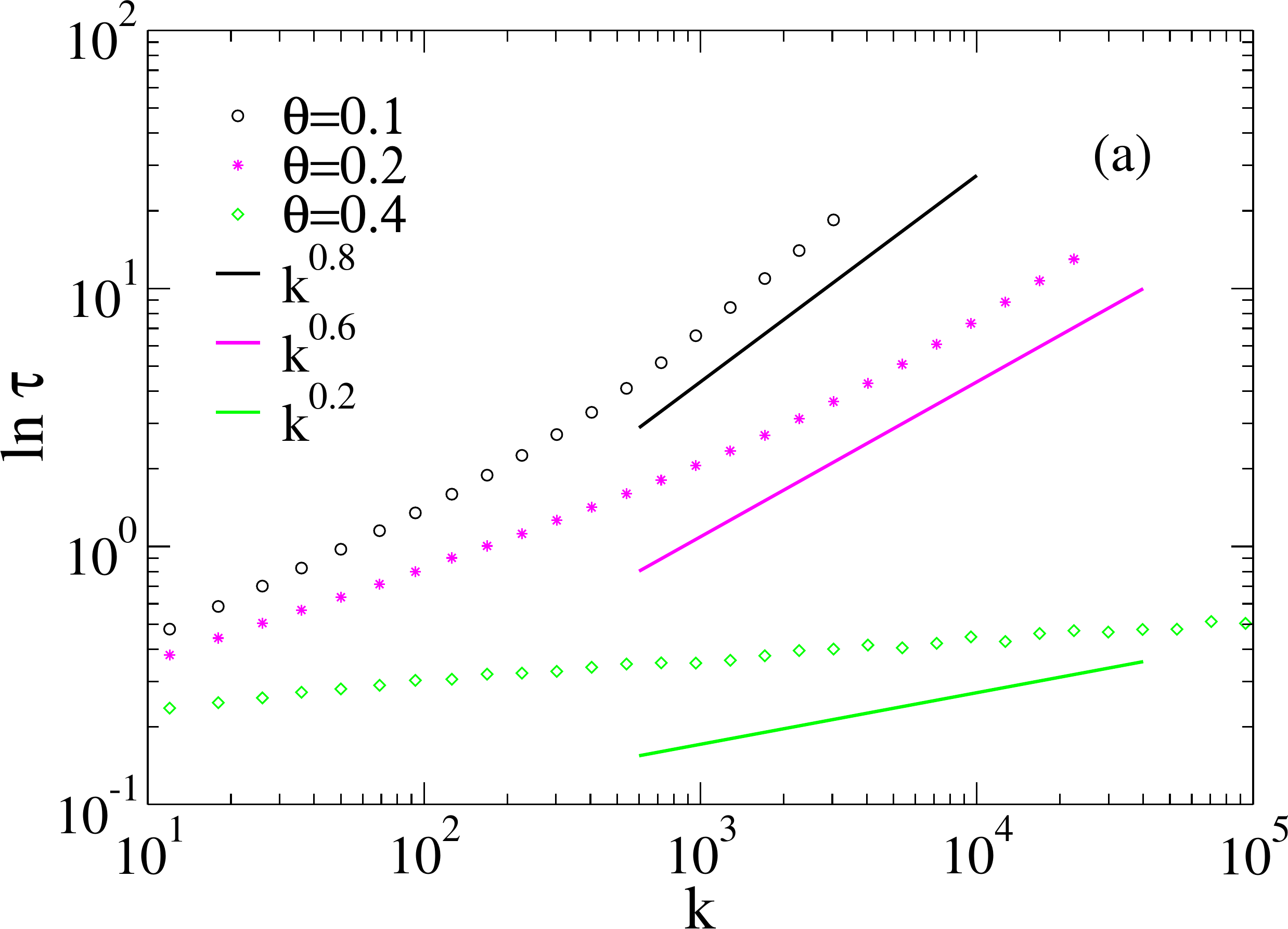}
 \includegraphics[width=0.8\linewidth]{\FigPath/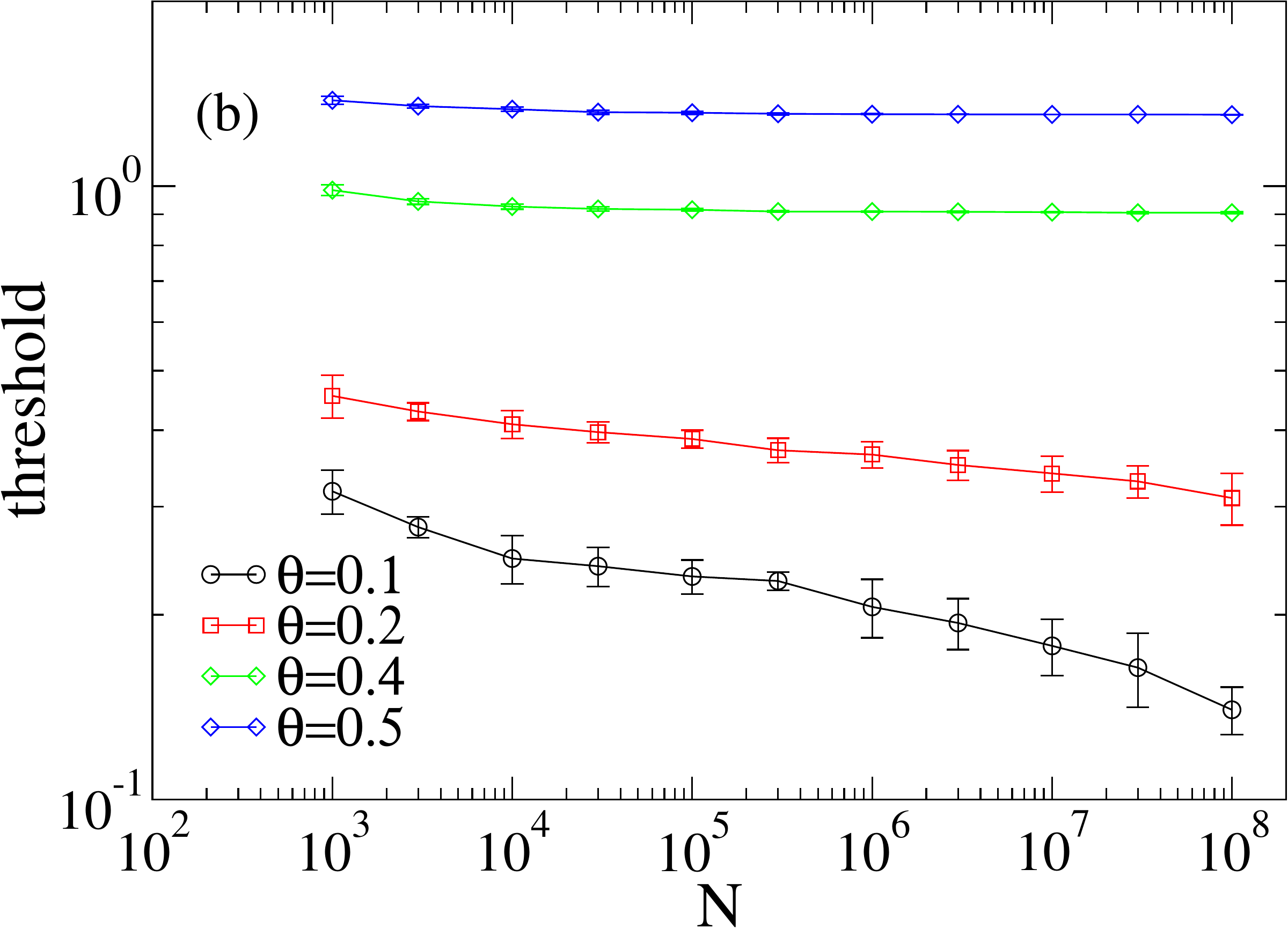}
 \caption{(color online) (a) Lifespan for KJI model on star graphs with
   $\lambda \av{k^{1-\theta}}/\av{k}=0.2$ confirming the stretched
   exponential asymptotic behavior expected for $\theta<1/2$.  (b)
   Epidemic activation thresholds for the KJI model with different
   values of $\theta$ and $\gamma = 3.5$.}
 \label{fig:KJI_th}
\end{figure}

Applying Eq.~(\ref{eq:1}) to the present model leads to
$\lbkin{k} = \lbkout{k} = \lambda
\lrangle{k^{1-\theta}}/(\av{k}k^{\theta})$, which translates, from
Eq.~\eqref{eq:taukSIS}, to a hub recovery time
$\tau^\mathrm{rec, KJI}_k \sim \exp(\mathrm{const}\cdot k^{1-2\theta})$
that is finite for $\theta>1/2$ and diverges as a stretched exponential
for $\theta<1/2$. These results are backed up by numerical simulations
of the KJI model on star graphs, see Fig.~\ref{fig:KJI_th}.

The mutual infection time of hubs for $\theta<1/2$ scales similarly as
in the SIS dynamics
$\tau^{\mathrm{inf,KJI}}\sim
N^{\frac{\gamma-3}{\gamma-1}b(\bar{\lambda})}$ for $\gamma>3$ and
$\tau^{\mathrm{inf,KJI}}\sim\mathrm{const}$ for $\gamma<3$,
the only difference being in the factor
$\bar{\lambda}=\lambda \lrangle{k^{-\theta}}
\lrangle{k^{1-\theta}}/\lrangle{k}$.  For $\theta \ge 1/2$, we have
$1/\lbkout{k}$ and thus $\tau^{\mathrm{inf,KJI}}$ diverging as in the CP
case.  Thus, in the case $\gamma>3$, where HMF and QMF predictions
disagree markedly for $\theta<1/2$, we obtain that, for $\theta<1/2$,
the transition should be driven by a hub activation mechanism, since in
this region $\tau^\mathrm{rec} \gg \tau^\mathrm{inf}$, and thus should
correspond to a vanishing threshold, qualitatively in agreement with
QMF, which indeed predicts a threshold
$\lambda_c^\mathrm{QMF,KJI} \sim k_\mathrm{max}^{\theta - 1/2}$, see
Appendix~\ref{app:KJI}.  For $\theta>1/2$, on the other hand, the hub
lifetime is finite, compatible with a collectively activated transition,
and corresponding to a finite threshold, in agreement with both HMF and
QMF theories. These predictions are verified in Fig.~\ref{fig:KJI_th} by
means of numerical simulations of the KJI model on SF networks using the
quasi-stationary (QS) method~\cite{cpquenched,DeOliveira05}, estimating
the effective threshold for each network size as the position of the
main peak shown by the susceptibility
$\chi=N(\lrangle{\rho^2}-\lrangle{\rho})/\lrangle{\rho}$ (see Appendix
\ref{app:simu} for simulation details).  For values of $\theta$ close to
$1/2$, however, long crossovers are observed in the threshold, in
analogy with the behavior observed in the lifespan of stars. Indeed,
after a crossover that can be very long as $\theta$ approaches 1/2, the
epidemic lifetime of KJI model on star graphs of increasing size reaches
the asymptotic regime of a stretched exponential, see
Fig.~\ref{fig:KJI_th}.  These crossovers, also observed in the numerical
estimate of the QMF threshold (see Appendix \ref{app:KJI}), explain the
apparently constant threshold observed in Fig~\ref{fig:KJI_th} for
$\theta = 0.4$. Equivalent scenarios can be drawn for $\gamma<3$ with
critical values of $\theta$ smaller than 1/2.

\subsection{SIRS model}

We finally consider the SIRS model, an extension of the SIS model, with
the same $\lambda_{k, k'}$ constant, but with a finite waning
immunity. Application of standard mean field theories (see
Appendix~\ref{app:SIRS}) leads to exactly the same result as the SIS
model, independently of the waning immunity $\alpha$, i.e
$\lambda_c^\mathrm{HMF, SIRS} = \av{k}/\av{k^2}$ and
$\lambda_c^\mathrm{QMF, SIRS} = 1/\Lambda_m$. So, we face the same
situation of the SIS and KJI models, with two contradictory predictions
in SF networks for $\gamma>3$

In the SIS model, from Eq.~(\ref{eq:taukSIRS}), and given that
$\lbkin{k} = \lbkout{k} = \lambda$, we obtain
\begin{equation}
  \tau_k^\mathrm{rec, SIRS} \sim k^{\alpha/\beta},
  \label{eq:6}
\end{equation}
that is, an algebraic increase of the hub recovery time with degree,
modulated by the exponents $\alpha$ and $\beta$.  This analytic
prediction is confirmed in numerical simulations of the SIRS model using
the QS method (see Appendix~\ref{app:simu} for
details) in Fig.~\ref{fig:tauk}(a). The agreement observed is expected
for small $\lambda$, since only a few leaves are infected in each step
and thus neglecting the recovered leaves at the end of each step becomes
a good approximation.

\begin{figure}[t]
  \centering
  \includegraphics[width=0.96\linewidth]{\FigPath/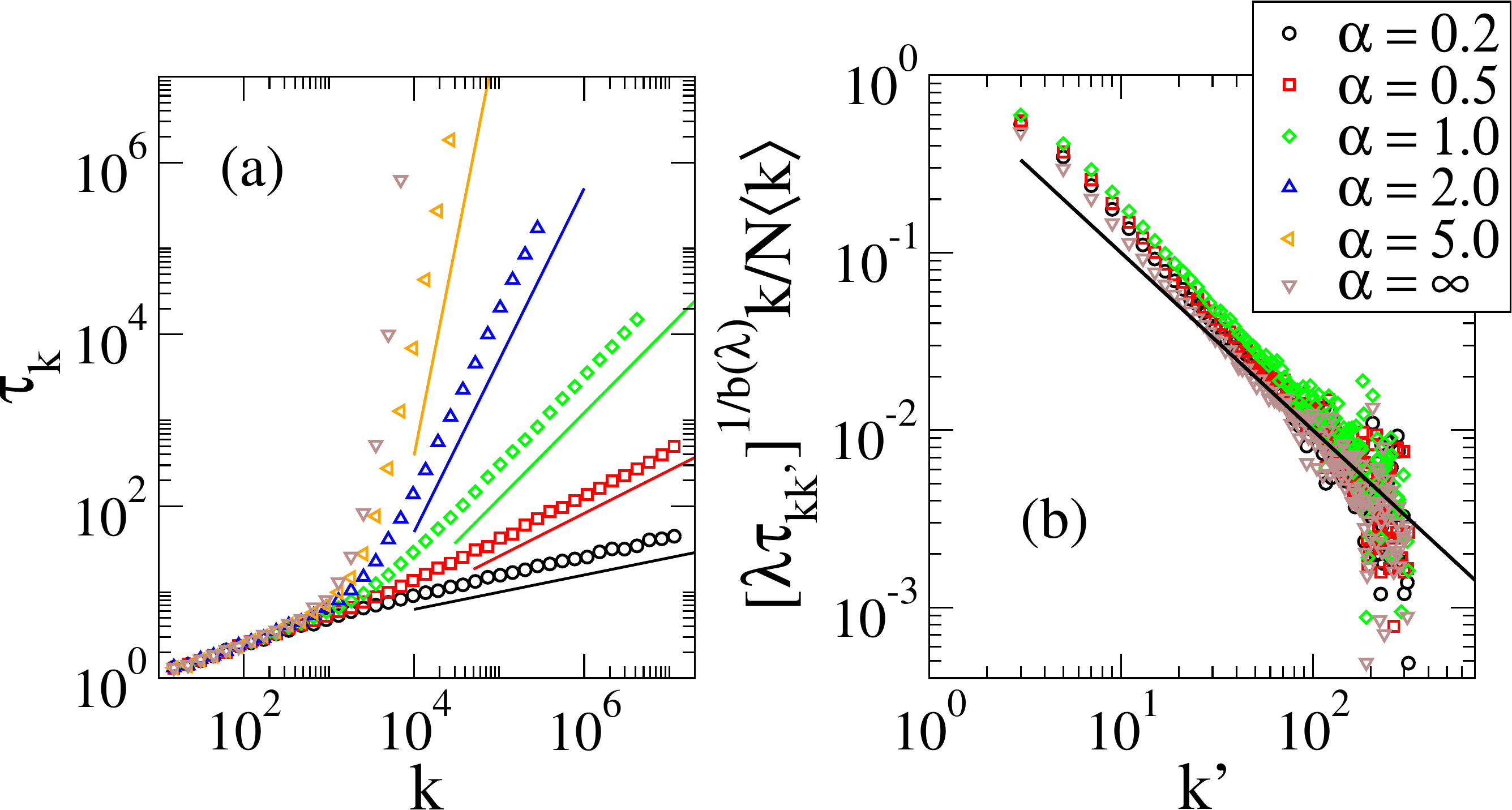}\\\vspace{0.2cm}
  \includegraphics[width=0.9\linewidth]{\FigPath/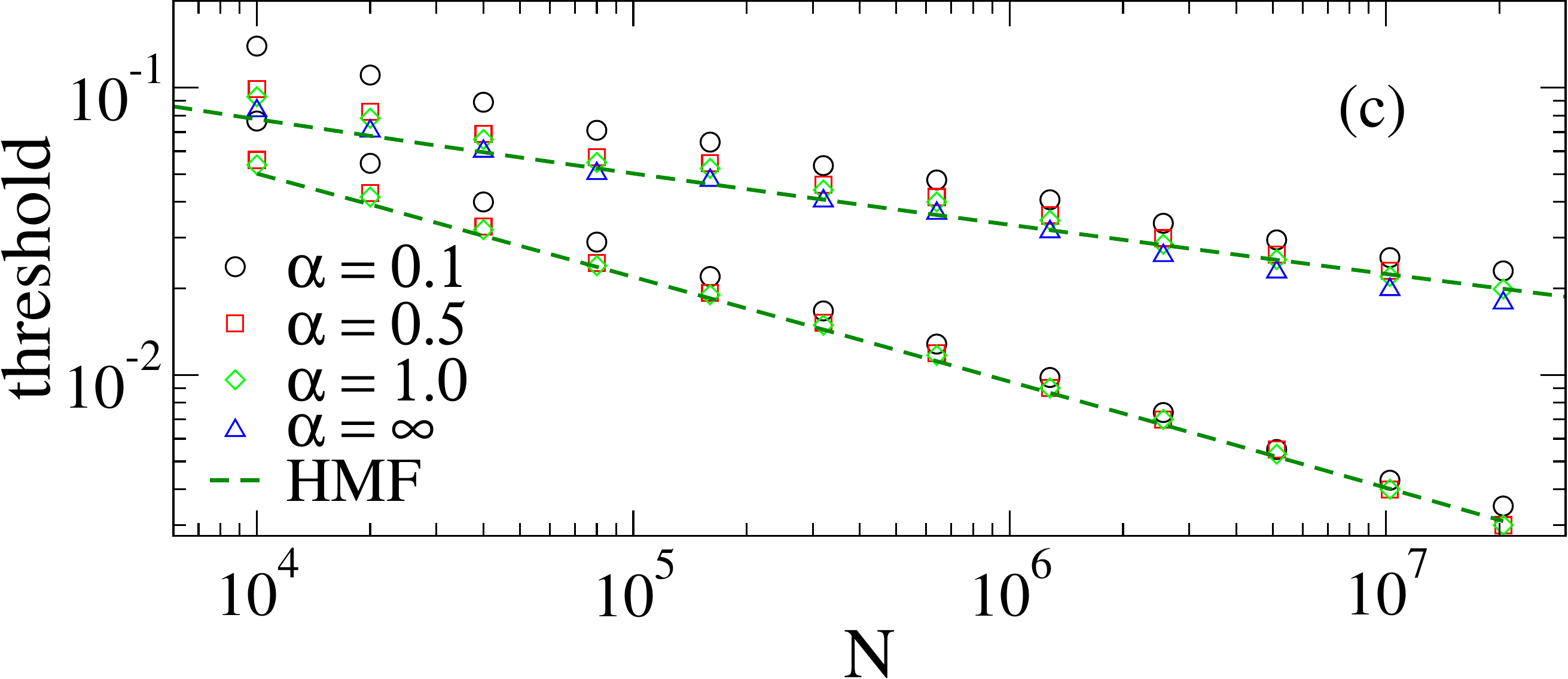}
  \caption{(color online) (a) Lifespan for SIRS/SIS dynamics on star
    graphs with $k$ leaves for $\lambda=0.05$, $\beta=1$ and different
    values of $\alpha$. 
    Solid lines are power laws $\tau_k\sim k^{\alpha/\beta}$. (b)
    Infection times of vertices of degree $k'$ in a network with
    $N=10^5$ vertices where the epidemics starts in a vertex of degree
    $k=50$ which is never cured. The solid line represents the
    theoretical value of Eq.~(\ref{eq:bound}).  (c) Epidemic thresholds
    against network size for different immunization times $1/\alpha$,
    using SF networks with degree exponents $\gamma=2.2$ (bottom curves)
    and $\gamma=2.7$ (top curves). Dashed lines correspond to
    $\lambda_c^\mathrm{QMF}=1 / \Lambda_m$, where $\Lambda_m$ is the
    largest eigenvalue of the adjacency matrix.}
  \label{fig:tauk}
\end{figure}

Application of Eq.~(\ref{eq:bound}) turns out the same result as in the
SIS model, i.e. a finite infection time for $\gamma<3$ and an algebraic
increase for $\gamma>3$. This result, which is independent of $\alpha$,
is numerically confirmed in Fig.~\ref{fig:tauk}(b). At this respect, it
is interesting to notice that the basic hypothesis used in our analysis,
see Sec.~\ref{sec:hub-lifetime}, considering only the epidemic routes
where the infected vertices always transmit the infection to its right
neighbor before it becomes recovered or
susceptible~\cite{boguna2013nature}, is more precise for longer
immunization periods (small $\alpha$) since the multiple infection of a
vertex in this path occurs rarely, implying the bound is an estimate of
$\tau^{\mathrm{inf}}_{k,k'}$ for SIRS as good as or better than that for
SIS.

Combining the previous results, we observe that for $\gamma<3$, the hub
recovery time is always larger than the hub infection time, the same
situation observed on the SIS mode, with the only difference of the
sharper (exponential) increase of $\tau^{\mathrm{rec}}_{k}$ in the SIS
case. The scenario that we expect in this range of $\gamma$ values is
thus a SIS-like transition, with the hub reinfection mechanism at work
and a vanishing epidemic threshold, in qualitative agreement with QMF.
In order to check the prediction, we have performed numerical
simulations of the SIRS model on SF networks using the QS method (see
Appendix~\ref{app:simu}) for details).  In Fig.~\ref{fig:tauk}(c) we
show that, for $\gamma<3$, even a very small value of $\alpha$ leads to
a scaling of the epidemic threshold against networks sizes in very good
agreement with the QMF prediction.

For $\gamma>3$, on the other hand, the situation is more complex and the
threshold finite-size behavior depends on $\alpha$.  From the divergence
of the the maximum degree in Eq.~(\ref{eq:5}), we obtain
$\tau^{\mathrm{inf,SIRS}}\sim N^{\frac{\gamma-3}{\gamma-1}b(\lambda)}$
and $\tau^\mathrm{rec,SIRS}\sim N^{\frac{\alpha/\beta}{\gamma-1}}$, that
is, algebraic increases with network size in both cases. A SIS-like
regime (hub reinfection triggering epidemics where the threshold
decreases with $N$) is expected whenever
$1/\beta \ll \tau^{\mathrm{inf,SIRS}}\ll \tau^\mathrm{rec,SIRS}$, which
corresponds to $b(\lambda)<\tfrac{\alpha}{\beta(\gamma-3)} \ln \kappa$
or, equivalently,
\begin{equation}
\lambda > \beta\vartheta (\alpha,\gamma) \equiv
\beta/\left[\kappa^{\alpha/[\beta(\gamma-1)]} -1 \right].
\end{equation}
Unless $\alpha$ is sufficiently small and/or $\gamma$ is sufficiently
large, this inequality is violated, and the hub lifetime is smaller than
the hub infection time. This indicates that the hub activation scenario
is not viable, and in analogy with the contact process, it hits towards
a finite threshold.  However, for sufficiently small
$\vartheta(\alpha,\gamma)$, we can observe a region of $\lambda$ values
for which the hub activation mechanism is at work, leading to an
effective threshold decreasing with $N$.  A sufficient condition to
observe this effective SIS-like behavior is
$\beta \vartheta (\alpha,\gamma)<\lambda_c^\mathrm{SIS}(N)$, the
effective SIS threshold in the network of size
$N$~\cite{Ferreira12,boguna2013nature,mata2014multiple}, since if the
SIS dynamics cannot activate hubs in a network, SIRS dynamics cannot
either, due to the suppressing effect of immunity. Assuming a scaling
$\lambda_c^\mathrm{SIS}(N) \sim k_\mathrm{max}^{-\mu}$ where $\mu=1/2$
for the QMF theory~\cite{Castellano10}, this SIS-like behavior should be
observed for network sizes
$N \ll N_c \equiv [\beta \vartheta(\alpha, \gamma)]^{(\gamma-2)/\mu}$,
crossing over to a constant threshold for $N \gg N_c$.

\begin{figure}[t]
 \centering
 \includegraphics[width=0.99\linewidth]{\FigPath/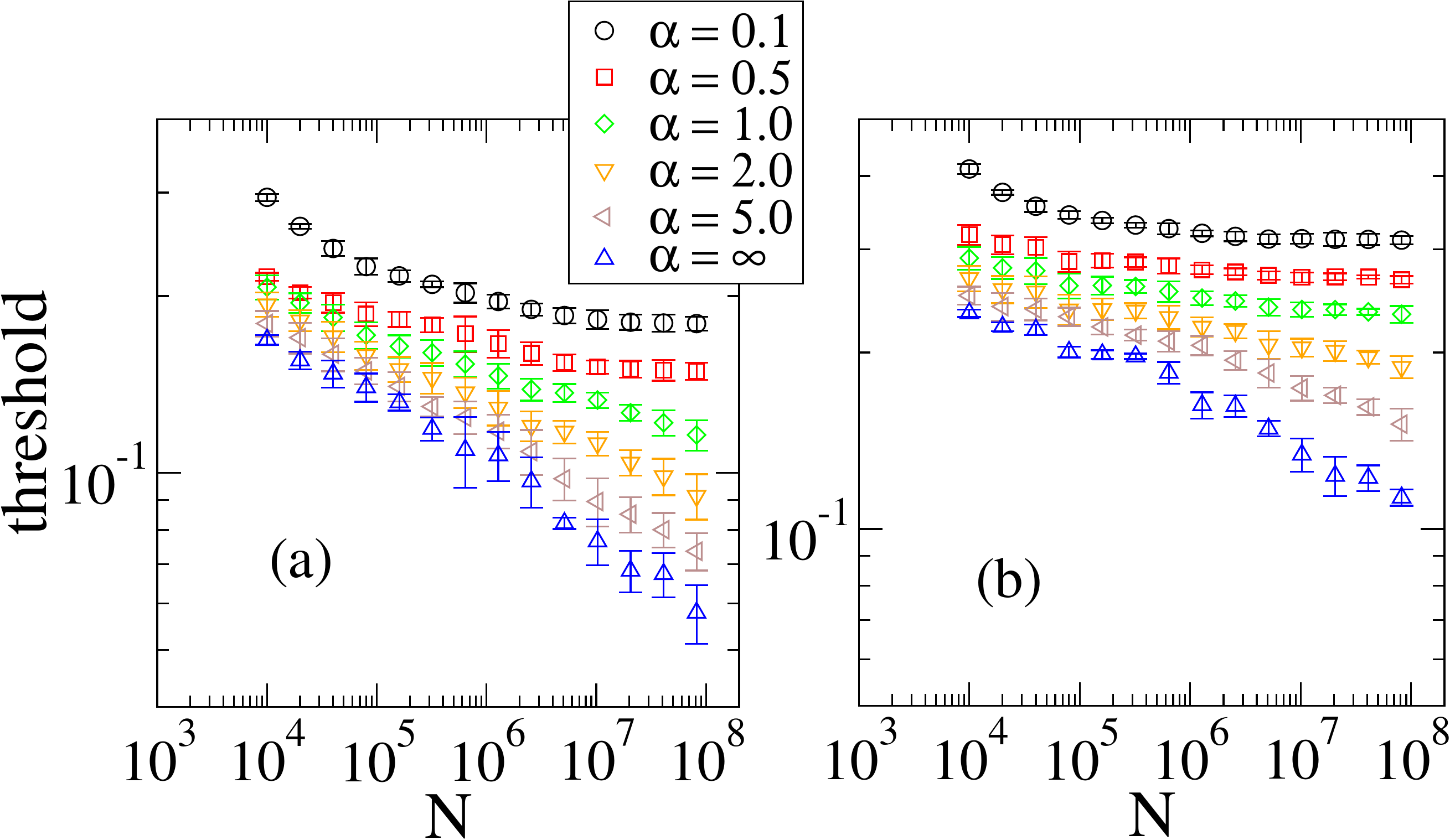}\\
 \caption{(color online) Epidemic threshold for SIRS model against
   network size for UCM networks with distinct degree exponents (a) $\gamma=3.5$
   and (b) $\gamma=4.0$ with minimal degree $k_0=3$. This SIS limit ($\alpha=\infty$) is
   also included for comparison.} 
 \label{fig:ther}
\end{figure}

Numeric thresholds for $\gamma>3$ are shown in Fig.~\ref{fig:ther}. 
As we can see, for sufficiently small $\alpha$ (up to $1$ for $\gamma
= 4$), we can observe a constant threshold for large $N$. For larger
$\alpha$ values, the trend is still decreasing, being the constant
plateau located on system sizes larger than those available to our
computer resources.

\begin{figure*}[t]
 \includegraphics[width=0.52\linewidth]{\FigPath/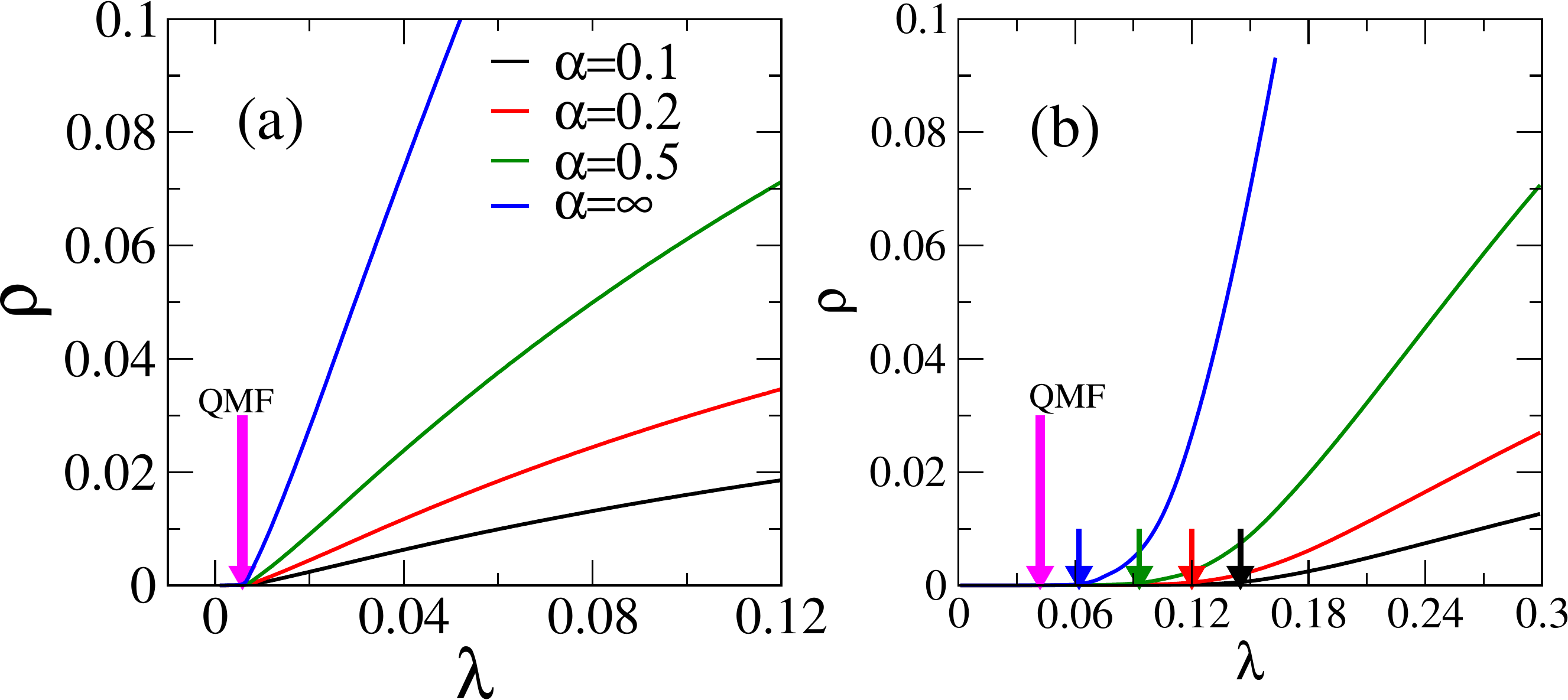}
 \includegraphics[width=0.25\linewidth]{\FigPath/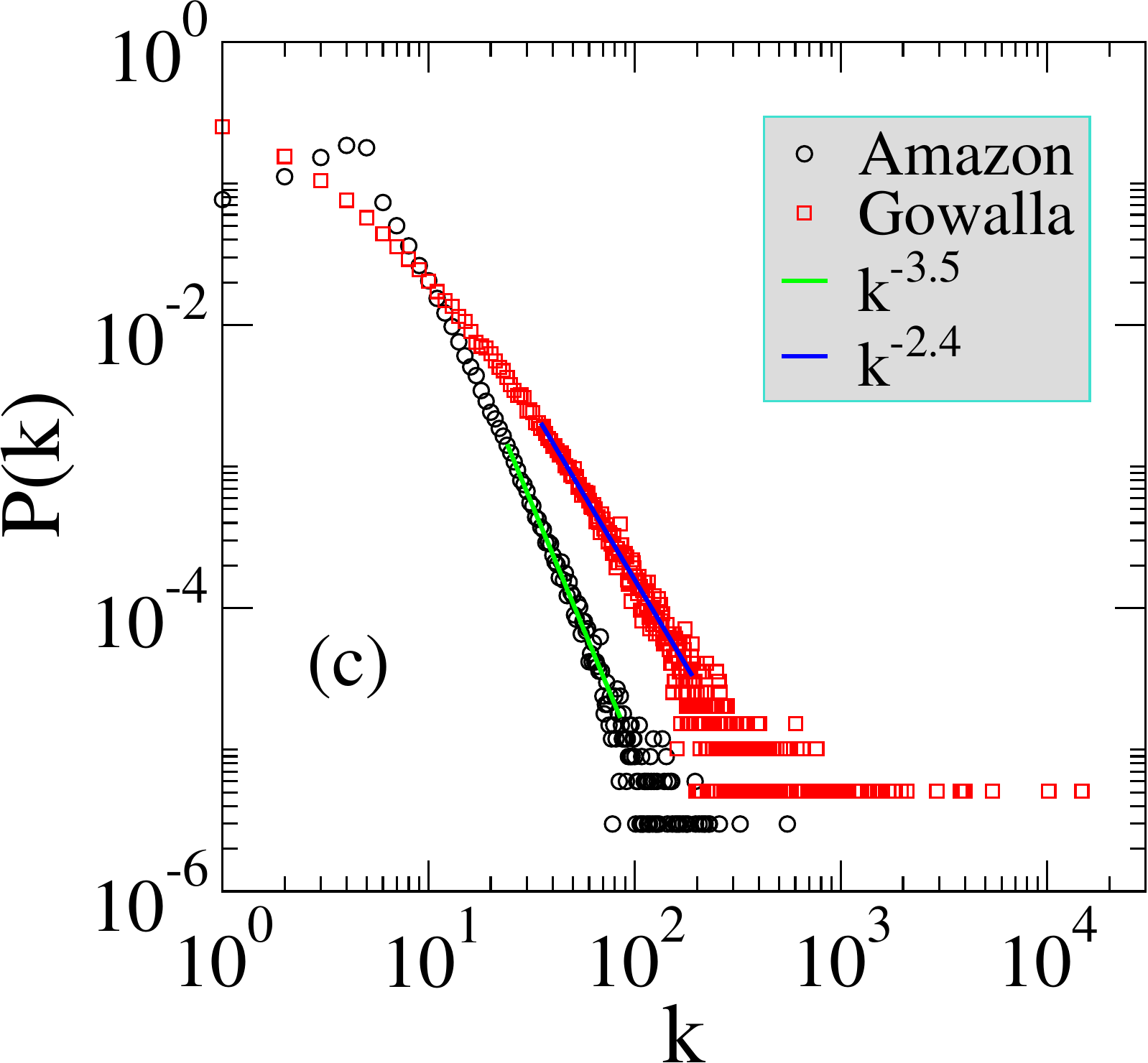}
 \caption{(Color online) {Simulation of SIS and SIRS dynamics on real networks.}
   Quasistationary density against infection rate in
   (a) Gowalla and (b) Amazon networks for different waning immunity
   rates. Arrows indicate the positions of the thresholds obtained via
   susceptibility method and QMF theory, see Appendix~\ref{app:SIRS}. (c)
   Degree distribution for Gowalla ($N=196591$, $\km=14730$,
   $\gamma=2.4$) and Amazon networks ($N=334863$, $\km=549$,
   $\gamma=3.5$). Solid lines are power-law regressions.}
 \label{fig:real}
\end{figure*}

\section{SIRS models on real networks}
\label{sec:real}

These results presented so far have been mainly checked on synthetic
uncorrelated networks. They can however be extended to real correlated
networks, characterized by conditional probability $P(k' | k)$ with a
non-trivial dependence on $k$~\cite{Dorogovtsev:2002}.  Focusing on the
SIRS model, in Figures \ref{fig:real}(a) and \ref{fig:real}(b) we
present numerical simulations on two SF real networks. We consider in
particular the location-based social network Gowalla~\cite{gowalla} and
the product co-purchasing network in Amazon website~\cite{amazon},
possessing degree exponents smaller and larger than $3$, respectively,
Figure~\ref{fig:real}(c).  According QMF theory, the thresholds for
these networks are equal to the inverse of the largest eigenvalue of
their adjacency matrix. By means of a numerical diagonalization, we
obtain the values $\lambda_c^\mathrm{QMF, Gowalla}=0.0059$ and
$\lambda_c^\mathrm{QMF, Amazon}=0.042$. Our simulations show, in the
case of the Gowalla network, with a degree exponent
$\gamma \simeq 2.4 < 3$, that numerically estimated thresholds are
essentially independent of $\alpha$ and very close to the QMF
prediction. This behavior is consistent with the theoretical expectation
of a hub activated dynamics, to be observed in the regime $\gamma<3$,
see Fig.~\ref{fig:real}(a). On the other hand, for the Amazon network,
with degree exponent $\gamma \simeq 3.5 >3$, we observe effective
thresholds that diminish with increasing $\alpha$, approaching the QMF
prediction for $\alpha\rightarrow\infty$.  This behavior is again in
agreement with the prediction for the SIRS model in SF networks with
$\gamma>3$, that indicates a finite threshold for small $\alpha$ values,
opposite to the QMF prediction of threshold independence.

\section{Summary}
\label{sec:conclu}

The determination of the properties of the epidemic transition in models
of disease propagation in highly heterogeneous networks in a crucial
topic in network science, to which a large research effort has been
recently devoted. Among others, one of the main questions that remain to
be answered in this field is what are the conditions under which a given
epidemic model leads to a vanishing or a finite threshold, and how the
properties of the epidemic transition can be best described from a
theoretical point of view. 

In the present paper, building on an extension of the theory of
Ref.~\cite{boguna2013nature}, we have proposed a general criterion to
discern the nature of thresholds in epidemic models. The criterion in
based in the comparison time scales of hub recovery (lifespan) and hub
reinfection. When the lifespan is larger than the infection time,
dynamics is triggered by a hub activation process, akin to the SIS
dynamics: Hubs survive for very long times, and are able to reinfect
each other, in such a way as to establish a long-lived endemic
state. This hub activation scenario leads to a vanishing threshold when
the lifespan is diverging in the thermodynamic limit. In this case, QMF
theories are expected to be qualitatively correct. The reason underlying
the effectiveness of QMF theories lies in the fact that they take into
account the full topological structure of the network, and are dominated
by the effects of the largest hubs. On the other hand, for a lifespan
smaller than the infection time scale, hub activation cannot be
sustained and possible epidemic phase transitions should be the result
of a collective activation process, leading to a standard phase
transition at a finite threshold, as in the case of the CP.  In this
second scenario, HMF theories should be qualitatively correct, due to
the fact the they work on the annealed network approximation, in which
every node can interact with every other with a degree dependent
probability \cite{Dorogovtsev08}.

To check the validity of the criterion, we have investigated a generic
epidemic model with spontaneous recovering, waning immunity, and edge
degree dependent infection rates on scale-free networks, for which we
can compute analytic expressions for the hub recovery and infection time
scales. The model has as particular cases the fundamental epidemic
models, as the SIS, SIRS and contact processes (CP) as well as more
complex ones as the generalized SIS model of Ref.~\cite{Karsai}, which
we have considered. After exemplifying our framework with the known
cases of the SIS and CP models, we present an extended discussion of the
SIRS model, an extension of the SIS model with waning immunity. While
previous analytic approaches (HMF and QMF theories) predict for the SIRS
model the same behavior than the SIS model, the main result from out
criterion is that we are able to show that, instead, the effect of even
a small amount of waning immunity is able to restore a finite threshold
(albeit affect by possible strong finite-size effects) in scale-free
networks with degree exponent $\gamma>3$, at odds with the QMF theory
valid for SIS in this regime, and in agreement with HMF. Our predictions
are corroborated by means of numerical simulation on synthetic
uncorrelated scale-free networks, as well as on real correlated
networks.

The proposed criterion represents a step forward in our understanding of
the properties of the epidemic transition in epidemic modeling, and thus
opens the path to study more general and realistic models. In this
sense, its application to more complex models is straightforward, only
possible hampered by technical difficulties in extracting analytic
expressions for the hub lifetime and infection time scales. These time
scales can, however, be numerically estimated from direct simulations of
epidemics on star networks, as we have shown in the examples
presented here.

\begin{acknowledgments}
  This work was partially supported by the Brazilian agencies CAPES, CNPq,
  and FAPEMIG. R.P.-S. acknowledges financial support from the Spanish
  MINECO, under Project No. FIS2013-47282-C2-2, and ICREA Academia,
  funded by the Generalitat de Catalunya. R.P.-S is special visiting
  researcher in the program \textit{Ci\^encia sem Fronteiras} - CAPES
  under project No.  88881.030375/2013-01.
\end{acknowledgments}

\appendix

\section{Mean field theory of the KJI model on networks}
\label{app:KJI}

In the Karsai, Juh\'{a}sz and Igl\'{o}i (KJI) model~\cite{Karsai}, an
edge transmits the infection from a vertex $j$ to vertex $i$ at a
weighted rate $\lambda_{ij}=\lambda A_{ij}/(k_i k_j)^\theta$, where
$\theta$ is a control parameter\footnote{A constant factor in the
  original definition was absorbed in $\lambda$.}. The HMF theory for
the KJI model is set in terms of the probability $I_k$ that a node of
degree $k$ is infected, while it is susceptible with probability
$1- I_k$. The rate equation for this quantity can be simply written as
\cite{Karsai,Dorogovtsev08,barratbook}
\begin{equation}
\frac{d I_k}{d t} = -I_k+ k(1-I_k)\sum_{k'}\frac{\lambda}{(k k')^\theta}
I_{k'} P(k'|k),
\label{eq:diKJI}
\end{equation}
where $P(k'|k)$ is the probability that an edge from a node of degree
$k$ points to a node of degree $k'$ \cite{alexei}.  The threshold is
obtained by linearizing Eq.~(\ref{eq:diKJI}) around the fixed point
$I_k =0$, which yields  for uncorrelated networks with
$P(k'|k) = k' P(k')/\av{k}$ \cite{Dorogovtsev:2002}
\begin{equation}
 \frac{d I_k}{d t} =\sum_{k'}L_{k k'}I_{k'},
\end{equation}
with a Jacobian
\begin{equation}
L_{k k'} = -\delta_{k k'}+\lambda (k k')^{1-\theta}P(k)/\lrangle{k}.
\end{equation}
The absorbing state $I_k =0$ loses stability when the largest eigenvalue
of $L_{k k'}$ is null. We thus obtain a threshold for the active state
with the form
\begin{equation}
\lambda_c^\mathrm{HMF} = \frac{\lrangle{k}}{\lrangle{k^{2(1-\theta)}}}.
\label{eq:hmfKJI}
\end{equation}
As we see from this equation, $\theta=0$ leads to the epidemic threshold
of the SIS model, depending on $\gamma$, while $\theta=1/2$ leads to
the threshold of the CP model, independent of the network structure. For
general values of $\theta$, a finite HMF threshold is expected for
$\gamma > 3 - 2\theta$, while it is null for $\gamma < 3 - 2\theta$.

From the point of view of QMF theory, based on the microscopic
probability $I_i$ that node $i$ is infected, the relevant rate equation
can be written as \cite{Newman10}
\begin{equation}
\frac{d I_i}{d t} = -I_i+\lambda (1-I_i)\sum_{j}I_{j} \frac{A_{
		ij}}{(k_i k_j)^\theta},
\label{eq:dIiKJI}
\end{equation}
where $A_{ij}$ is the adjacency matrix \cite{Newman10} with value
$A_{ij} = 1$ if nodes $i$ and $j$ are connected, and zero otherwise, and
we consider the normalization condition $ I_i+S_i=1$. After
linearization, stability analysis leads to a threshold inversely
proportional to the largest eigenvalue $\Lambda_m^D$ of the matrix
\begin{equation}
D_{ij}=\frac{A_{ij}}{(k_i k_j)^\theta}.
\label{eq:qmfKJI} 
\end{equation}

\begin{figure}[t]
\centering
\includegraphics[width=0.8\linewidth]{\FigPath/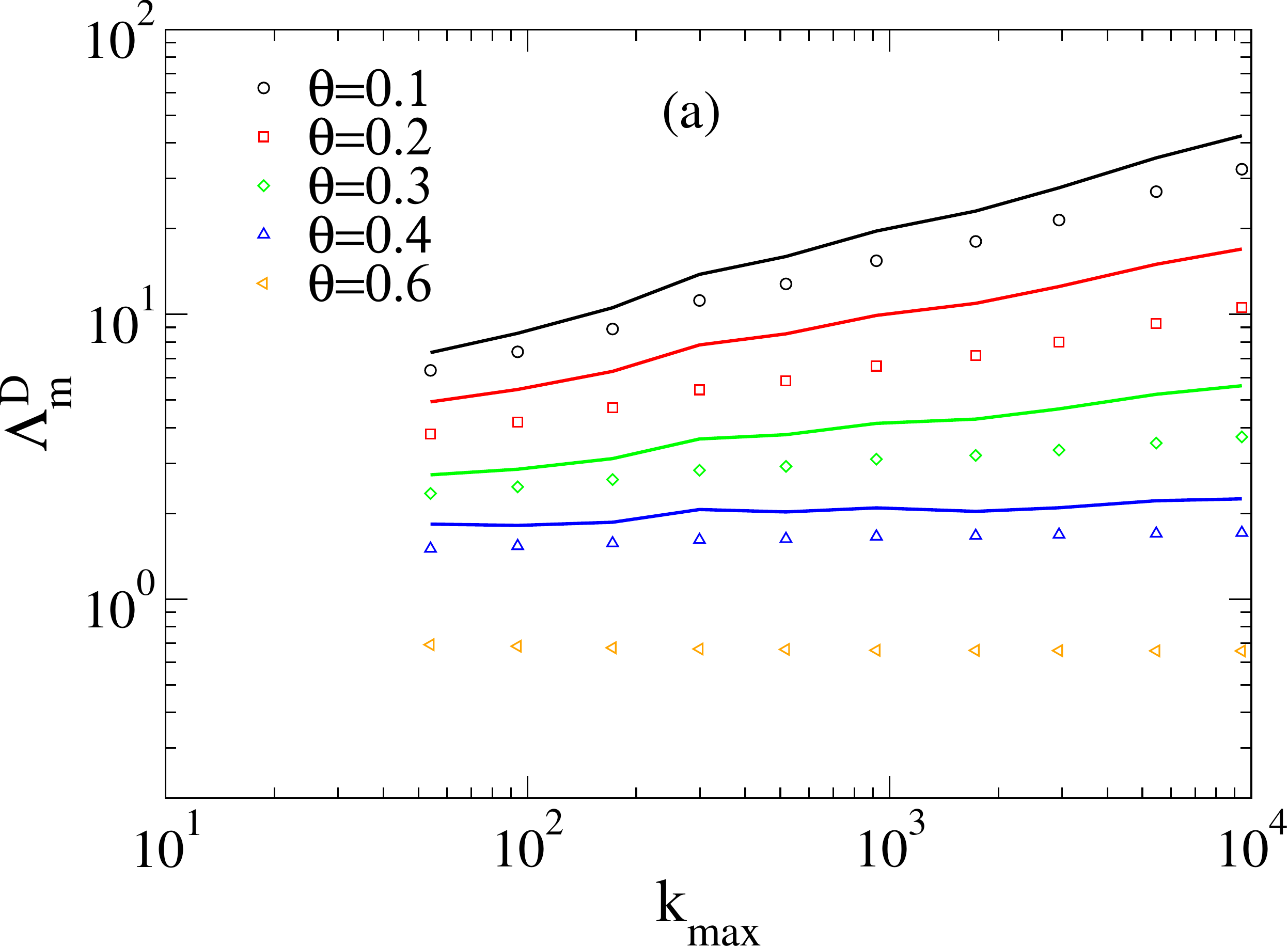} ~~
\includegraphics[width=0.8\linewidth]{\FigPath/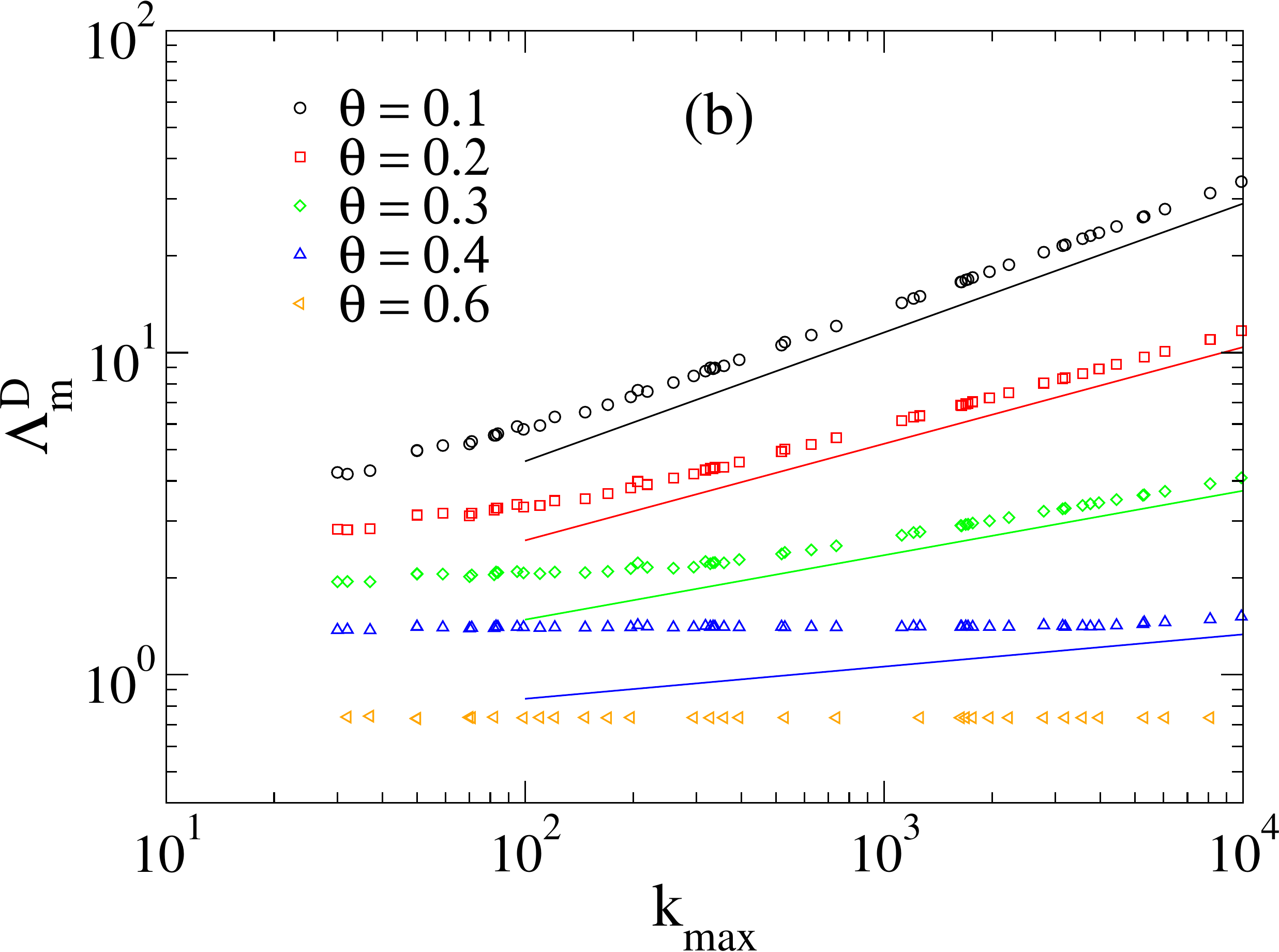}
\caption{(color online) Scatter plot of the largest eigenvalue
  $\Lambda_m^D$ of the matrix $D_{ij}=A_{ij}/(k_i k_j)^\theta$ against
  the degree of the most connected vertex $k_\mathrm{max}$ for UCM
  networks with exponents (a) $\gamma=2.7$  and (b) $\gamma=3.5$
  using minimal degree $k_0=3$ and structural cutoff
  $k_c\sim N^{1/2}$~\cite{Catanzaro05}. The results were computed for 5
  independent networks for $\gamma=3.5$ and 1 for $\gamma=2.7$ (largest
  degree fluctuates little). Sizes from $N=10^3$, $3\times 10^3$,
  $10^4$, $3\times 10^4$, $10^5$, $3\times 10^5$, $10^6$,
  $3\times 10^6$, $10^7$, $3\times 10^7$, and $10^8$ were used.  In the
  bottom panel, solid lines are power laws
  $k_\mathrm{max}^{\frac{1}{2}-\theta}$.  In the top panel, solid lines
  are proportional to $\Lambda_m/k_{max}^\theta$, with $\Lambda_m$ being the
  numerically estimated largest eigenvalue of the adjacency matrix
  $A_{ij}$ for $\theta>1/2$, and a constant for
  $\theta>1/2$.}
\label{fig:Lmax}
\end{figure}  

No general analytical expression is available for the largest eigenvalue
of this matrix, so we have proceeded to determine it numerically in SF
networks generated using the uncorrelated configuration model (UCM)
\cite{Catanzaro05}, see Fig.~\ref{fig:Lmax}. We find that
$\Lambda_m^D\sim \Lambda_m/k_{max}^\theta$ for $\theta<1/2$, where
$\Lambda_m$ is the largest eigenvalue of the adjacency matrix $A_{ij}$
for $\theta>1/2$, and a constant for $\theta>1/2$.  For $\gamma>3$,
since $\Lambda_m\sim \sqrt{\km}$ \cite{Chung03}, we have
$\Lambda_m^D \sim k_{max}^{\frac{1}{2}-\theta}$ for $\theta < 1/2$ and
$\Lambda_m^D \sim \mathrm{const.}$ for $\theta \geq 1/2$.  The behavior
of the largest eigenvalue in this case has strong finite size effects
close to $\theta = 1/2$. These size effects can be observed in the
crossover from a flat region to the scaling regime
$k_{max}^{\frac{1}{2}-\theta}$ for $\theta<1/2$, crossover that takes
place at larger values of $k_\mathrm{max}$ when $\theta$ approaches
$1/2$.

These observations indicate that, for $\theta <1/2$, a zero threshold is
obtained in the thermodynamic limit, independently of $\gamma$, while a
finite threshold should occur for $\theta>1/2$.

\section{Mean field theories for the SIRS model on networks}
\label{app:SIRS}

In the HMF theory, the densities of infected, recovered and susceptible
vertices of degree $k$, are represented by $I_k$, $R_k$, and $S_k$,
respectively, and obey the normalization condition $I_k+R_k+S_k=1$. The
HMF dynamic equations, setting $\beta=1$, are given by
\cite{ForestFireSatorras09,Dorogovtsev08,barratbook}
\begin{equation}
 \frac{d I_k}{d t} = -I_k+\lambda kS_k \sum_{k'}I_{k'} P(k'|k),
\label{eq:dik}
\end{equation}
and 
\begin{equation}
  \frac{d R_k}{d t} = -\alpha R_k+I_k.
\label{eq:drk}
\end{equation}
To determine the threshold where an active state becomes stable, we
perform a linear stability analysis around the fixed point $I_k=R_k=0$,
corresponding to the absorbing state. Since we are interested in long
times, a quasi-static approximation \cite{michelediffusion}
$\frac{d R_k}{d t}\approx 0$ is used to
obtain $I_k=\alpha R_k$, which is inserted in Eq.~\eqref{eq:dik} to
result in a  linearized equation
with  Jacobian 
\begin{equation}
L_{k k'}= -\delta_{k k'}+\lambda k P(k'|k).
\label{eq:jacohmf}
\end{equation}
The absorbing state loses stability when the largest eigenvalue of
$L_{k k'}$ is null.  Thus, for uncorrelated networks with
$P(k'|k)=k'P(k')/\lrangle{k}$, we easily obtain
that the infected state is stable for~\cite{ForestFireSatorras09}
\begin{equation}
 \lambda > \lambda_c^\mathrm{HMF}=\frac{\lrangle{k}}{\lrangle{k^2}}.
 \label{eq:lbchmf}
\end{equation}

In the QMF theory, the process is defined in terms of the probabilities that
a vertex $i$ is infected, $I_i$, recovered, $R_i$, or susceptible,
$S_i$, which fulfill the equations
\begin{equation}
 \frac{d I_i}{d t} = -I_i+\lambda S_i \sum_{j}I_{j} A_{ ij},
 \label{eq:dIi}
\end{equation}
and 
\begin{equation}
\frac{d R_i}{d t} = -\alpha R_i+I_i,
\label{eq:dRi}
\end{equation}
where $A_{ij}$ is the adjacency matrix \cite{Newman10}, and we consider
the normalization condition $ I_i+R_i+S_i=1$.

Applying a quasi-static approximation to Eqs.~\eqref{eq:dIi} and
\eqref{eq:dRi}, we obtain the linearized equation
\begin{equation}
\frac{d I_i}{d t} =\sum_j L_{ij}I_j,
\end{equation}
with the Jacobian $L_{ij}=-\delta_{ij}+\lambda A_{ij}$, implying that the
threshold is given by~\cite{mata2013pair}.
\begin{equation}
 \lambda_c^\mathrm{QMF}=\frac{1}{\Lambda_m},
\end{equation}
where $\Lambda_m$ is the largest eigenvalue of the adjacency matrix.

\section{Simulation methods}
\label{app:simu}

SIRS simulations were implemented for $\beta=1$ adapting the
simulation scheme of Refs.~\cite{Ferreira12,mata2014multiple}: At each
time step, the number of infected nodes $N_i$, the number of edges
emanating from them $N_k$, and the number of recovered vertices $N_r$,
are computed and time is incremented by $dt = 1/(N_i+\lambda N_k +
\alpha N_r)$. With probability $N_i/(N_i+\lambda N_k + \alpha N_r)$
one infected node is selected at random and becomes recovered.  With
probability $\alpha N_r/(N_i+\lambda N_k + \alpha N_r)$, a recovered
vertex is select and turned to susceptible. Finally, with probability
$\lambda N_k/(N_i+\lambda N_k+\alpha N_r)$, an infection attempt is
performed in two steps: (i) An infected vertex $j$ is selected with
probability proportional to its degree.  (ii) A nearest neighbor of
$j$ is selected with equal chance and, if susceptible, is infected. If
the chosen neighbor is infected or recovered nothing happens and
simulation runs to the next time step. The numbers of infected and
recovered nodes and links emanating from the former are updated
accordingly, and the whole process is iterated.

KJI simulations were performed generalizing the previous algorithm as
follows.  For each node of the network we calculate the weight
\begin{equation}
w_i=\sum_{j}A_{ij}(k_i k_j)^{-\theta},
\end{equation} 
proportional to the total infection rate transmitted by all edges of
$i$. Then, at each time step, the number of infected nodes $N_i$ and
the sum of the weights $w_i$ over all infected nodes $S_w$ are
computed and time is incremented by $dt = 1/(N_i+\lambda S_w)$. With
probability $N_i/(N_i+\lambda S_w)$ one infected node is selected at
random and becomes susceptible.  With probability $\lambda
S_w/(N_i+\lambda S_w)$, an infection attempt is performed in two
steps: (i) An infected vertex $i$ is selected with probability
proportional to its weight $w_i$. (ii) A nearest neighbor of $i$,
namely $j$, is selected with probability proportional to
$k_j^{-\theta}$ and, if susceptible, it is infected.

\begin{figure}[t]
\centering
\includegraphics[width=0.8\linewidth]{\FigPath/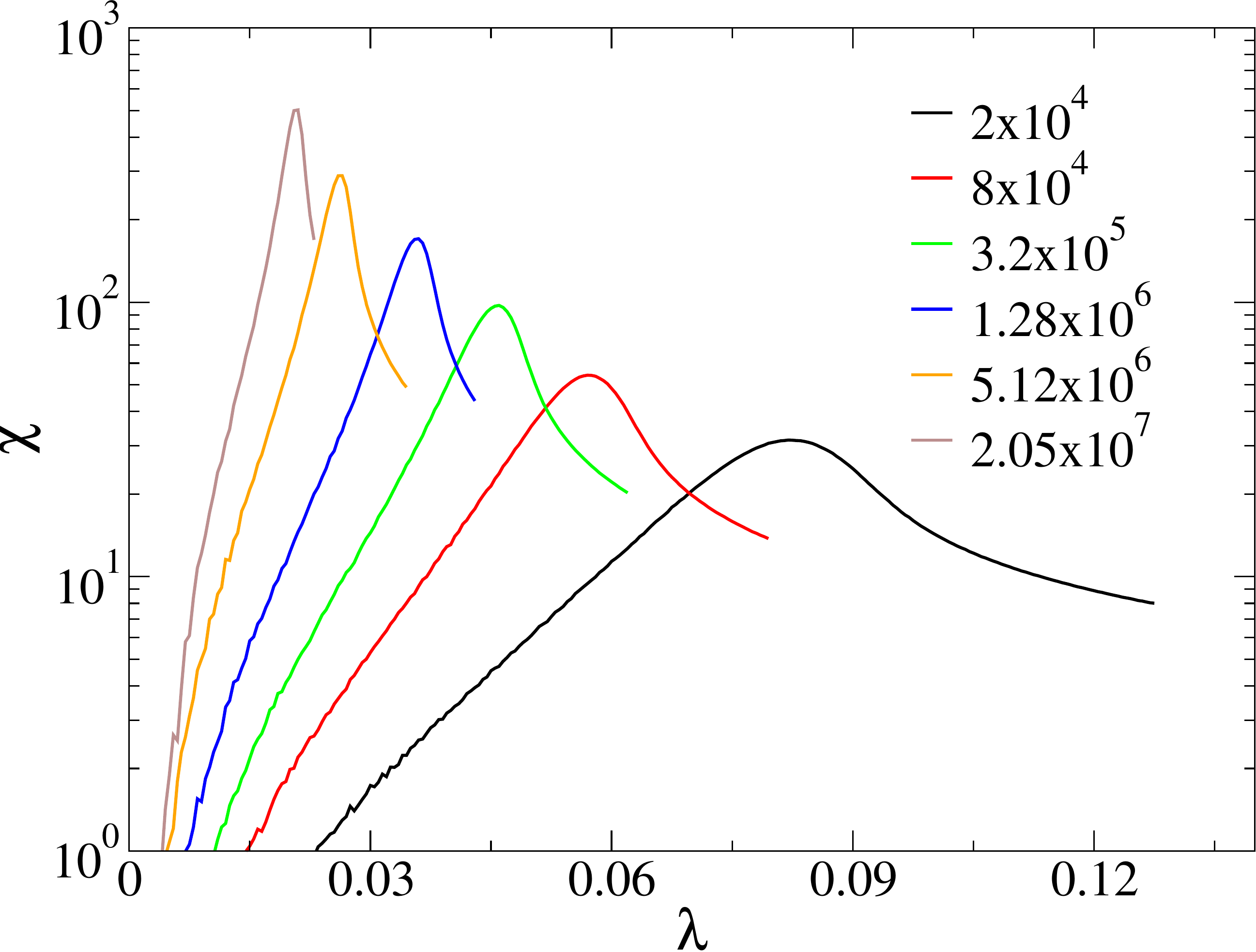}
\caption{Susceptibility against infection rate for SIRS dynamics with
  $\alpha=1.0$.
  We used UCM networks of different sizes (quoted in the labels) with
  $k_0=3$ and $\gamma=2.7$.}
\label{fig:susg2p7a}
\end{figure}

The simulations were performed using the quasi-stationary (QS) method
~\cite{DeOliveira05,annealed2011} which permits to overcome the
difficulties intrinsic to the simulations of finite systems with
absorbing states. In the QS method, every time the system visits an
absorbing state it jumps to an active configuration previously visited
during the simulation. This task is achieved building and constantly
updating a list containing $M=70$ configurations. The update is done by
randomly picking up a stored configuration and replacing it by the
current one with probability $p_r\Delta t$.  We fixed
$p_r\simeq 10^{-2}$ since no dependence on this parameters was detected
for a wide range of simulation parameters. After a relaxation time
$t_r$, averages are computed over a time $t_{av}$. Typically, a QS state
is reached at times $t \gtrsim 10^4$ for QS simulations of dynamical
processes on complex networks. Therefore, $t_r=10^5$ was used in all
simulation. On the other hand, averaging times from $10^6$ to $10^8$
were used, the larger the average time the smaller the infection rate.

During the averaging time, the QS probability $\bar{P}(n)$ that the
system has $n$ infected vertices is computed. All stationary
quantities of interest can be derived from $\bar{P}(n)$. Here, we will
investigate the density of infected vertices
$\rho=\sum_nn\bar{P}(n)/N$, the lifespan $\tau=1/\bar{P}(1)$ and the
susceptibility defined as
$\chi=N(\av{\rho^2}-\av{\rho}^2)/\av{\rho}$~\cite{Ferreira12}. This
susceptibility has a diverging peak at the transition to an absorbing
state on complex networks~\cite{Ferreira12,Mata14,RonanEPJB} and has
been successfully used to determine the thresholds for epidemic
models~\cite{Ferreira12,mata2013pair,Lee2013,wang2014}.

Simulations were done on SF networks with $N$ vertices and degree
distribution $P(k)\sim k^{-\gamma}$ generated with the uncorrelated
configuration model (UCM)~\cite{Catanzaro05} with minimum degree $k_0=3$
and structural upper cutoff $k_{max}=N^{1/2}$, which guarantees absence
of degree correlations in the networks generated, that are, therefore,
suitable for comparisons with the HMF theory where this simplification
was adopted. Averages were computed using more than $20$ different
network samples.

The determination of the threshold, estimated as the peak of
susceptibility, is shown in Fig.~\ref{fig:susg2p7a}.


%

\end{document}